\documentclass[journal=jpcbfk,manuscript=article]{achemso}

\usepackage{chemformula} 
\usepackage[T1]{fontenc} 
\usepackage{bm}
\usepackage{amsmath}
\usepackage{color}
\usepackage{xcolor}
\usepackage{multirow}

\graphicspath{{figures/}}

\author{Nathan London}
\affiliation[Cornell University]{Department of Chemistry, Cornell University, Ithaca, NY}
\alsoaffiliation{Current address: Division of Energy, Matter and Systems, School of Science and Engineering, University of Missouri-Kansas City, Kansas City, MO}
\author{Siyu Bu}
\affiliation[Cornell University]{Department of Chemistry, Cornell University, Ithaca, NY}
\author{Britta Johnson}
\affiliation[Cornell University]{Department of Chemistry, Cornell University, Ithaca, NY}
\alsoaffiliation{Current address: Physical Science Division, Pacific Northwest National Laboratory}
\author{Nandini Ananth}
\email{ananth@cornell.edu}
\affiliation[Cornell University]{Department of Chemistry, Cornell University, Ithaca, NY}

\title{Mean-Field Ring Polymer Rates Using a Population Dividing Surface}

\keywords{American Chemical Society, \LaTeX}

\begin{document}




\begin{abstract}
Mean-field Ring Polymer Molecular Dynamics (MF-RPMD) offers 
a computationally efficient method for the simulation of 
reaction rates in multi-level systems. Previous work has established that,
to model a nonadiabatic state-to-state reaction accurately, the 
dividing surface must be chosen to explicitly sample kinked ring 
polymer configurations where at least one bead is in a different 
electronic state than the others. Building on this, we introduce 
a population difference coordinate and a kink-constrained dividing surface, and we test the accuracy of the resulting mean-field rate theory on a series of linear vibronic coupling model systems as well as spin-boson models. We demonstrate that this new MF-RPMD rate approach is efficient to implement and quantitatively accurate for models over a wide range of driving forces, coupling strengths, and temperatures.  
\end{abstract}

\section{Introduction} \label{sec:intro}

Nonadiabatic condensed phase reactions play a critical role in understanding reaction 
mechanisms for a range interesting systems including proton coupled electron transfer in biological systems and charge transfer 
in energetic materials.~\cite{augMarcus1985,julGray1996,maySkourtis2010,sepAzzouzi2018,
	janYarkony2012,octHammes-Schiffer2008}
Despite the development of numerous nonadiabatic dynamic methods for the characterization of 
reaction rates and mechanisms, atomistic simulations of multi-level systems that incorporate
nuclear quantum effects and vibronic couplings remain largely out of reach.

Perhaps the most ubiquitous and successful nonadiabatic rate theory is the Marcus theory for electron transfer  that assumes a classical solvent.~\cite{mayMarcus1956} 
Going beyond Marcus theory, Fermi's Golden Rule (FGR) rate theory incorporates nuclear quantum effects, and can be computed exactly for parabolic potentials. In certain regimes, FGR rates can be approximated using Wolynes rate theory,~\cite{decWolynes1987} and limitations of the Wolynes approach have been addressed by recent work, \textcolor{black}{extending the methods applicability to the Marcus inverted regime and improving its accuracy for 
high temperatures and anharmonic potentials.~\cite{marLawrence2018,decFang2019,febFang2020,octLawrence2020, octLawrence2020a}
Despite these advances,  these theories are confined to the weak perturbation limit where FGR is applicable, although 
an interpolation strategy has been proposed to move between the golden-rule limit theories 
and adiabatic rate theory approaches.~\cite{sepLawrence2019} }

 Dynamical methods developed to study the 
 mechanisms of nonadiabatic processes and to compute rate constants 
 include exact methods like MultiConfiguration Time-Dependent Hartree (MCTDH)~\cite{janBeck2000,augHan2022} 
 real-time path-integral based approaches,~\cite{junMakri1998,decChatterjee2019,julMakri2020}
 and Heirarchical Equations of Motion,~\cite{janTanimura1989,aprTanimura1991,octXing2022} 
 in addition to approximate methods like Redfield theory,~\cite{janRedfield1957,febKimura2016} Fewest Switches 
 Surface Hopping and other mixed quantum-classical dynamics methods,~\cite{julTully1990,mayKapral2006,octJain2015,mayCrespo-Otero2018, sepMiller2016, febCotton2013,junWarburton2022} 
and Semiclassical theories.~\cite{febCao1997,julCao1995,octRichardson2015a,octRichardson2015,mayLee2016,augAnanth2007,marChurch2018,decLiu2021} 
 Unfortunately, the majority of these approaches 
 are not suitable for large scale system simulations, with a notable exception being systems that can be modeled by spin-boson Hamiltonians where transitions between electronic states are mediated by coupling to a nuclear bath. Approximate imaginary-time path integral methods like 
 Centroid Molecular Dynamics (CMD)~\cite{aprCao1994, julJang1999} and 
 Ring Polymer Molecular Dynamics (RPMD)~\cite{augCraig2004} show great promise in moving beyond the spin-boson models to the more complex bilinear coupling Hamiltonian models and even atomistic simulations.~\cite{febCraig2005,janLawrence2020,novNovikov2018,febHele2013,augAlthorpe2013,mayHele2016,aprHabershon2013,augMenzeleev2011,augKenion2016,octKretchmer2016,augLiao2002,janPavese1999,octLoose2022,janPaesani2010,novPaesani2008,julReichman2000} . 
 Both CMD and RPMD conserve the quantum Boltzmann distribution 
 and capture nuclear quantum effects like tunneling using an ensemble 
 of classical trajectories in an extended phase space, making them efficient 
 for the simulation of systems with many degrees of freedom. 
Nonadiabatic extensions of CMD have been explored in the context of the spin-boson model,~\cite{augLiao2002} while RPMD has been used for atomistic simulations of nonadiabatic electron transfer and proton-coupled electron transfer~\cite{augMenzeleev2011,octKretchmer2016}
 However, it has been shown that RPMD cannot capture state-to-state dynamics when working with a two-level or, in general, $K$-level system Hamiltonian.~\cite{augMenzeleev2011} 
 This limitation has motivated the development of 
several multi-state imaginary-time path integral based methods including 
 mean-field (MF)-RPMD \cite{janSchwieters1998, Hele2011, junDuke2016}, 
 nonadiabatic RPMD \cite{julRichardson2013},
 kinetically constrained (KC)-RPMD \cite{febMenzeleev2014, junKretchmer2016,mayKretchmer2018}, coherent-state RPMD
 \cite{decChowdhury2017}, and mapping-variable (MV)-RPMD. \cite{decAnanth2010, sepAnanth2013, octDuke2015,
 decPierre2017} 
 \textcolor{black}{Other approaches have also been developed that combine RPMD with {\it adhoc} nonadiabatic dynamics
 methods like surface hopping.~\cite{octShushkov2012,marTao2018,febTao2019,decLawrence2019} }

 \textcolor{black}{
 Of these methods, only MV-RPMD and MF-RPMD retain the key RPMD feature of conserving the quantum Boltzmann distribution,
 and MF-RPMD is unique in that it does not introduce any explicit electronic variables, 
 instead describing nuclear dynamics on an effective state-averaged potential energy surface. This makes MF-RPMD 
 more approximate but significantly more efficient than MV-RPMD.} Initial efforts to compute MF-RPMD rates used the nuclear centroid reaction coordinate and defined the dividing surface as configurations where the two electronic states were degenerate.~\cite{Hele2011} It was shown that the resulting rate expression significantly overestimated nonadiabatic reaction rates.~\cite{Hele2011} Subsequent work has established that this error can be attributed to the use of a dividing surface that does not require the formation of kinked ring \textcolor{black}{polymer configurations} (where at least one bead is in a different electronic state than the others), an essential intermediate step in state-to-state reactions.~\cite{junKretchmer2016, junDuke2016} Further, it has been established that this error can be mitigated by introducing a doubly-constrained dividing surface that requires formation of kinked configurations, however the resulting MF-RPMD rate expression proved challenging to implement.~\cite{junDuke2016} 
 
Here we introduce an improved reaction coordinate that tracks the difference in electronic state populations, and a modified reactive flux definition that accounts for the formation of kinked ring polymer configurations.
We demonstrate that this new coordinate and reactive flux definition enable accurate calculations of nonadiabatic and adiabatic reaction rates for a series of bilinear coupling model systems, captures nonadiabatic electron transfer 
rates across a wide range of driving forces, and successfully 
incorporates nuclear quantum effects by reproducing FGR rates over a range of temperatures in models where the driving force places them in the Marcus normal regime. In the Marcus inverted regime, we show MF-RPMD reproduces Marcus theory rates rather than FGR rates. This failure to account for nuclear tunneling can be traced to the use of a Boltzmann weighted reaction coordinate that effectively limits the the ring polymer to sample only nuclear configurations where the two electronic states are degenerate. 

 \section{Methods} \label{sec:theory}

\subsection{Path Integrals for Multi-level Systems}
\begin{figure}[h]
     \centering
     \includegraphics[scale=1]{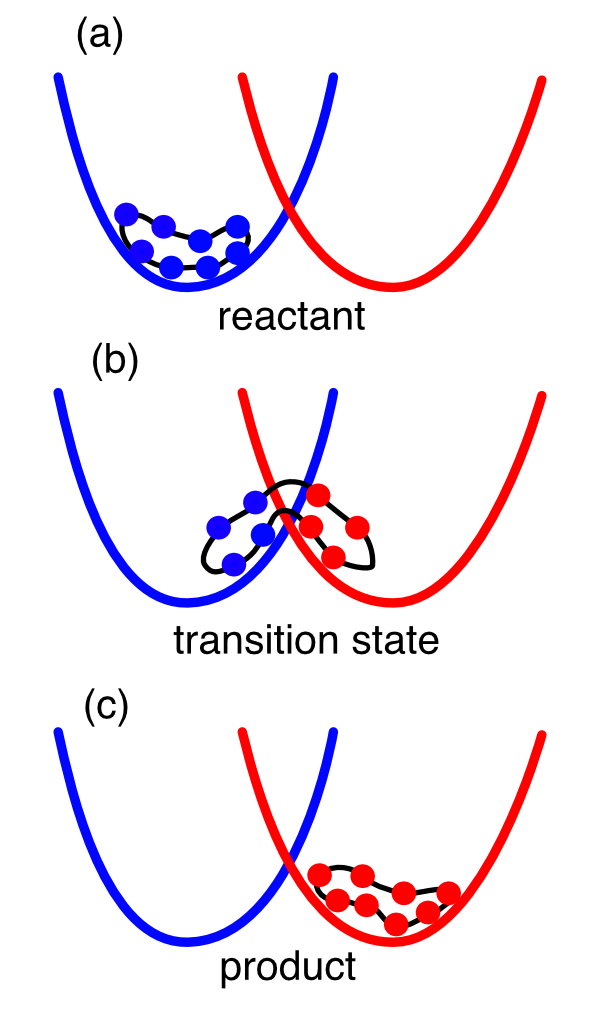}
 \caption{Cartoon showing the progress of a multi-level system reaction with state-space path integrals. In the reactant and product states, ring polymer configurations where all beads are in a single electronic state are the dominant contributions to the effective mean-field potental. At the transition
 state, when the system is moving from one state to the other, we expect to see a significant contribution from multi-electronic state or kinked ring polymer configurations.~\cite{aprChandler1981}}
	\label{fig:mf-rxn}
 \end{figure}

For a general $K$-level system with $d$ nuclear degrees of freedom, the diabatic Hamiltonian is 
\begin{equation}
    \hat{H} = \sum_{j=1}^d \frac{ \hat{P}_j^2  }{ 2 M_j} + \sum_{n,m = 1}^K | n \rangle V_{nm}(\hat{R}) \langle m |,
\end{equation}
where $\hat{R}$ and $\hat{P}$ are $d$-dimensional nuclear position and momentum vector operators, respectively,
$M_j$ is the nuclear mass, $\{|n\rangle\}$ is the set of diabatic states, and $V_{nm}(\hat{R})$ is the diabatic
potential energy operator. 
The quantum canonical partition function is obtained by evaluating the trace in the basis of nuclear positions and electronic states, and discretized via repeated insertion of $N$ copies of identity in the form of completeness relations to obtain
\begin{eqnarray}
    \nonumber
    Z &=& \mathrm{Tr}[e^{-\beta \hat{H}}] \\ 
		&=& \int d\{R_\alpha\} \sum_{\{n_\alpha\}=1}^K \prod_{\alpha=1}^N 
    \langle R_\alpha,n_\alpha | e^{-\beta_N \hat H} | R_{\alpha+1},n_{\alpha+1} \rangle,
    \label{eq:partfn}
\end{eqnarray}
where $\beta_N = 1/(N k_B T)$, T is temperature, 
N is the number of imaginary time slices (or beads), 
and $R_\alpha$, $n_\alpha$ refer to the nuclear position 
and electronic state, respectively, of the $\alpha^\text{th}$ bead. 
In Eq.~\ref{eq:partfn}, the shorthand, 
$\int d\left\{ R_\alpha \right\}=\int dR_1 \int dR_2..\int dR_N$
and $\sum_{ \{n_\alpha\}=1 }=\sum_{n_1}\sum_{n_2} \ldots \sum_{n_N}$,
is used to represent the multi-dimensional
integral over nuclear coordinates and summation over electronic states. 
The matrix elements in Eq.~\ref{eq:partfn} are evaluated using the Trotter (high temperature) approximation
to obtain \cite{Chandler1987,augTrotter1959},
\begin{eqnarray} \label{eq:PF_PIMD}
    Z \propto \lim_{N \rightarrow \infty} \int \{ dR_{\alpha} \} e^{ -\beta_N V_N(\{R_{\alpha}\})} \text{Tr}[ \Gamma ],
\end{eqnarray}
where
\begin{equation} \label{eq:rp_potential}
    V_N = \sum_{j=1}^d \sum_{\alpha=1}^N \left[\frac{M_j}{2 \beta_N^2} (R_{j,\alpha} - R_{j,\alpha+1})^2 \right],
\end{equation}
\begin{equation}\label{eq:gamma}
    \Gamma = \prod_{\alpha=1}^N \mathbf{M}(R_{\alpha}),
\end{equation}
and $\mathbf{M}$ is the $K \times K$-dimensional matrix
\begin{equation}\label{eq:m_matrix}
    M_{nm}(R_{\alpha}) = 
    \begin{cases}
    e^{-\beta_N V_{nn}(R_{\alpha})} & n=m \\
    -\beta_N V_{nm}(R_{\alpha}) e^{-\beta_N V_{nn}(R_{\alpha})}. & n\neq m
    \end{cases}
\end{equation}
The trace in Eq.~\ref{eq:PF_PIMD} can be moved into the exponential,
 and $N$ normalized nuclear momentum gaussians can be introduced 
 to express the canonical partition function as a phase space integral,
\begin{equation}
    Z \propto \lim_{N \rightarrow \infty} \int \{ dR_{\alpha} \} \int \{ dP_{\alpha} \} e^{ -\beta_N H_N( \{R_{\alpha}\},\{ P_{\alpha} \} ) },
\end{equation}
where the Mean-Field Ring Polymer Hamiltonian is 
\begin{equation}\label{eq:mfrpmd_hamiltonian}
    H_{N} = \sum_{j=1}^{d} \sum_{\alpha=1}^{N} \left[ \frac{M_j}{2 \beta_N^2} (R_{j,\alpha}-R_{j,\alpha+1})^2 + \frac{P_{j,\alpha}^2}{2 M_j} \right] - \frac{1}{\beta_N} \ln{(\text{Tr}[\Gamma])}.
\end{equation}
In Eq.~\ref{eq:mfrpmd_hamiltonian}, the Hamiltonian is a function of nuclear phase 
space variables only, with an effective mean-field potential,
$V_\text{MF}=-\frac{1}{\beta_N} \ln{(\text{Tr}[\Gamma])}$, obtained 
by tracing over all possible electronic state configurations for the ring polymer. 
During a state-to-state reaction, the contributions to the effective potential 
from ring polymer configurations where all the beads are in a single electronic 
state and configurations where the beads are in different electronic states will vary as the 
system moves from the reactant to product state as shown in Fig.~\ref{fig:mf-rxn}.

\subsection{The MF-RPMD Approximation to Reaction Rates}
The MF Hamiltonian in Eq.~\ref{eq:mfrpmd_hamiltonian} can be used 
to generate molecular dynamics trajectories that sample equilibrium configurations
of the $K$-level system at finite temperature as is done in standard 
Path Integral Molecular Dynamic simulations.
MF-RPMD approximates the Kubo-transformed real-time quantum correlation function 
by sampling trajectory initial conditions from an exact 
quantum canonical ensemble and time-evolving 
them under the mean-field Hamiltonian. 
By construction, MF-RPMD trajectories preserve the quantum Boltzmann distribution.

\textcolor{black}{
The MF-RPMD approximation to the flux-side correlation function, in keeping with the RPMD approximation,~\cite{julCraig2005} 
can be written as,
\begin{equation} \label{eq:rate_expression_general}
    C^\text{MF}_\text{fs}(t)= \frac{ \langle \delta( \xi_{0} - \xi^{\ddagger}) \dot{\xi_0} h(\xi_t - \xi^{\ddagger}) \rangle}
    {\langle h(\xi^{\ddagger} - \xi_0)\rangle},
\end{equation}
where $\langle\dots\rangle$ indicates a canonical ensemble average.
The reaction coordinate, 
$\xi$, in Eq.~\ref{eq:rate_expression_general} is a function of the 
ring polymer nuclear configurations $\{R\}$, with $\xi_0$ being its value
at time $t=0$, $\xi_t$ the value at any later time $t$,
and $\xi^{\ddagger}$ defines the dividing surface.
For reactions involving transitions between electronic states,
although there are no 
explicit electronic state variables in the MF-RPMD Hamiltonian, previous work suggests 
that it is essential that the definition of the reactive flux properly 
accounts for the formation of `kinked' ring polymer configurations as shown 
in Fig~\ref{fig:mf-rxn}.~\cite{aprChandler1981, febMenzeleev2014, junDuke2016} 
Unfortunately, for nonadiabatic processes, 
sampling the dividing surface $\xi=\xi^\ddag$ in Eq.~\ref{eq:rate_expression_general}
with the mean-field Hamiltonian yields an ensemble of 
ring polymer configurations where all beads are in a single electronic state 
rather than kinked configurations.
} 
The origin 
of this difficulty is well understood: the probability of kink formation is proportional to 
the square of the diabatic coupling, making this a rare event in the nonadiabatic regime where
coupling values are very small. In the language of the mean-field effective potential,
\begin{align}
    \label{eq:gamm_decomp}
    \Gamma=\Gamma_1 + \Gamma_2 + \textcolor{black}{\Gamma_\mathrm{k}},
\end{align}
where $\Gamma_i$, $i=1,2$, represent the potential contribution from configurations where all ring polymer beads are in the same electronic state,\textcolor{black}{
\begin{align}
\label{eq:state_gamm}
\Gamma_i = \prod_{\alpha=1}^N \left[ \mathbf{M}(R_{\alpha})\mathcal{P}_i \right],
\end{align}}
where $\mathcal{P}_i$ is the projection operator matrix corresponding to the $i^\text{th}$ electronic state,
and the matrix elements of $\mathbf{M}$ are defined in Eq.~\ref{eq:m_matrix}. 
The contribution to the effective potential from kinked ring polymer configurations, $\textcolor{black}{\Gamma_\mathrm{k}}$, has a coupling-square dependence that ensures that it is \emph{not} the dominant contribution to the $\Gamma$ function at any nuclear configuration. 

\textcolor{black}{
To 
sample \emph{only} kinked configurations at the dividing surface, 
it is necessary to introduce an {\it adhoc} modification to the dividing surface
that re-weights the sampling, 
\begin{align}
    \delta(\xi-\xi^\ddag)\rightarrow 
    \delta(\xi-\xi^\ddag)
    \frac{\text{Tr}\left[ \textcolor{black}{\Gamma_\mathrm{k}} \right]}{\text{Tr}\left[ \Gamma \right]}.
    \label{eq:new_ds}
\end{align}
Using the modified dividing surface, the MF-RPMD rate can be expressed as,
\begin{equation}\label{eq:kink-rate}
    k = \lim_{t \rightarrow t_\text{plateau}}\frac{ \int d\{R_{\alpha},P_\alpha\} e^{-\beta_N H_{N}^0}\text{Tr}
		\left[ \Gamma_\mathrm{k} \right] 
        \delta(\xi_0 - \xi^\ddag) \dot{\xi_0} h(\xi_t - \xi^\ddag)}
	{\int d\{R_{\alpha},P_\alpha\} e^{-\beta_N H_{N}^0}\text{Tr} \left[ \Gamma_{1} \right]},
\end{equation}
where $H_N^0$ is the free ring polymer Hamiltonian 
that includes only the harmonic spring term and the kinetic energy terms of Eq.~\ref{eq:mfrpmd_hamiltonian}
and the presence of $\color{black}{\Gamma_\mathrm{k}}$ ensures that the effective potential samples only kinked configurations. The time-evolved reaction coordinate, $\xi_t$, is obtained by integrating the classical equations of motion due to the MF-RPMD Hamiltonian in 
Eq.~\ref{eq:mfrpmd_hamiltonian}.
The denominator in Eq.~\ref{eq:kink-rate} is the reactant partition function, and is calculated by sampling configurations projected onto the reactant (here, state 1).}

We now introduce a new reaction coordinate, a bead-averaged population difference estimator that was previously 
employed in the context of MF-RP instanton calculations,~\cite{janSchwieters1998, marRanya2020} and is exact at equilibrium,
\begin{equation} \label{eq:deltaP}
	\Delta P = \frac{1}{N} \sum_{\alpha=1}^{N} \frac{e^{-\beta_{N}V_{22}(R_{\alpha})}
		-e^{-\beta_{N}V_{11}(R_{\alpha})}}{e^{-\beta_{N}V_{11}(R_{\alpha})}
		+e^{-\beta_{N}V_{22}(R_{\alpha})}}.
	\end{equation}
This coordinate takes negative values in the reactant region, 
positive values in the product region, and is
zero at the dividing surface. The corresponding MF-RPMD rate expression is then,
\begin{equation}\label{eq:mf-kink-rate}
    k = \lim_{t \rightarrow t_\text{plateau}}\frac{ \int d\{R_{\alpha},P_\alpha\} e^{-\beta_N H_{N}^0}\text{Tr}
		\left[ \Gamma_\mathrm{k} \right] 
\delta(\Delta P_0) \dot{\Delta P_0} h(\Delta P_t)}
	{\int d\{R_{\alpha},P_\alpha\} e^{-\beta_N H_{N}^0}\text{Tr} \left[ \Gamma_{1} \right]},
\end{equation}

\textcolor{black}{
The population coordinate and dividing surface introduced here are significantly different from the coordinate previously
introduced by one of us.~\cite{junDuke2016} In that work,
  the dividing surface was defined as nuclear configurations where half the ring polymer beads
  were in one state and the other half were in a different state. In addition, the reaction coordinate
  measured population differences but with the so-called Boltzmann estimator,~\cite{decPierre2017}
  is given by 
  \begin{align}
      \chi = \frac{\text{Tr}\left[
          \prod_{\alpha=1}^N {\bf M}(R_{\alpha})\left|2\right\rangle\left\langle 2\right|
          \right]
          - \text{Tr}\left[
          \prod_{\alpha=1}^N {\bf M}(R_{\alpha})\left|1\right\rangle\left\langle 1\right|
          \right]}
      {\text{Tr}\left[
          \prod_{\alpha=1}^N {\bf M}(R_{\alpha})\left|2\right\rangle\left\langle 2\right|
          \right]
          + \text{Tr}\left[
          \prod_{\alpha=1}^N {\bf M}(R_{\alpha})\left|1\right\rangle\left\langle 1\right|
          \right]}
      \label{eq:old_coord}
  \end{align}
  While both population estimators are exact at equilibrium in the mean-field path integral framework,
  the coordinate used here, $\Delta P$ as defined Eq.~\ref{eq:deltaP}, has been previously
  used to track changes in electronic state populations along the mean-field ring polymer
  instantons.~\cite{marRanya2020,janSchwieters1998} There are two significant advantages to
  the present $\Delta P$ coordinate. First, this coordinate is a bead-average population
  difference coordinate ensuring that the time zero correlation function correctly
  corresponds to a Kubo-transform real-time thermal correlation function. Second, $\Delta P$
  can be easily computed in large scale simulations where the donor and acceptor state
  potentials are known ($V_{11}$ and $V_{22}$), and does not require the matrix multiplication
  steps necessary to construct the $\chi$ coordinate. As a final point of comparison, we note
  that the velocity corresponding to $\chi$ was computed numerically using finite difference,~\cite{junDuke2016}
  whereas the analytic time derivative of $\Delta P$ is easily computed.
}

\textcolor{black}{
Our new rate expression defined in Eq.~\eqref{eq:mf-kink-rate} has some parallels with the rate formalism of KC-RPMD.~\cite{junKretchmer2016} Both methods emphasize the formation of kinked RP configurations to accurately described nonadiabatic reaction rates, but differ in the details. In KC-RPMD, the reaction coordinate is defined by an auxiliary dynamical variable that reports on the formation of kink-pairs; the introduction of this variable results in a modified Boltzmann distribution that the KC-RPMD Hamiltonian conserves. The MF-RPMD 
rate presented here does not introduce any additional dynamical variables, with our reaction coordinate depending only on the RP bead positions and the diabatic potential energy surfaces. Further, 
in KC-RPMD, the restriction that nonadiabatic transitions should occur at configurations where the diabatic states are near-degenerate is enforced by the use of a penalty function on kinked configurations in the effective potential in the form of a Gaussian of the scaled energy difference of the diabatic surfaces. In the present work, 
we ensure kink formation near the diabatic state crossing through our modified dividing surface. 
}

\subsection{Reference Rate Theories}

We compare nonadiabatic MF-RPMD rates computed here against 
Marcus theory and FGR rates.
The Marcus theory rate for a 
nonadiabatic process with a classical solvent is 
\cite{augMarcus1985},
\begin{equation}
k_{\text{MT}} = 
\frac{2 \pi}{\hbar} {| V_{nm}|}^2 
\sqrt{\frac{\beta}{4 \pi \lambda}} 
e^{-\beta \left( \lambda - \varepsilon \right)^2 / 4 \lambda},
\label{eq:kmt}
\end{equation}
where $\lambda$ is the reorganization energy,
$\varepsilon$ is the driving force, and $V_{nm}$ is
the diabatic coupling between the reactant and product 
electronic states.

FGR rate theory for a nonadiabatic reaction 
using the flux correlation formalism is \cite{novMiller1983},
\begin{equation}\label{eq:FGR}
	k_\mathrm{FGR} = \frac{1}{Z_0} \frac{\Delta^2}{\hbar^2}\int_{-\infty}^{\infty}dt\  C^{\tau}(t),
\end{equation}
where 
\begin{equation}\label{eq:fluxcor}
	C^{\tau}(t) = \mathrm{Tr} \left[ \mathrm{e}^{i \hat{H}_0 (t + i (\beta\hbar - \tau))/\hbar} 
		\mathrm{e}^{-i \hat{H}_1(t - i\tau)/\hbar} \right],
\end{equation}
	$Z_0 = \mathrm{Tr} \left[ \mathrm{e}^{-\beta\hat{H}_0} \right]$ is the reactant partition function. $\hat{H}_0$ and
		$\hat{H}_1$ are the Hamiltonians representing the reactant and product states respectively. $\tau$ can be any real
		number as the rate is independent of its choice \cite{novMiller1983} but is, typically, chosen to be in the range
		$[0,\beta\hbar]$. For the bilinear coupling model Hamiltonians employed here, the FGR
  rate can be obtained analytically;~\cite{marMattiat2018} 
  Details of this calculation are provided in the Supporting Information.

In the adiabatic limit, the electron transfer rate constant for a 
system with quantum solvent can be expressed using Kramers theory,
	\cite{janHenriksen2008l}
\begin{equation}\label{eq:KT}
	k_\mathrm{KT} = \left( \sqrt{1+\left(\frac{\gamma}{2\omega_\mathrm{b}}\right)^2} - \frac{\gamma}{2\omega_\mathrm{b}}
		\right) \frac{\omega_\mathrm{s}}{2\pi} \mathrm{e}^{-\beta G_\mathrm{cl}^{\ddagger}},
\end{equation}
where $\omega_\mathrm{s}$ is the frequency of the harmonic potential, $\omega_\mathrm{b}$ is the frequency at the barrier top, 
$\gamma=\eta/M_\mathrm{s}$, $\eta$ is the strength of the
	coupling to a dissipative bath, $M_\mathrm{s}$ is the solvent mass, and $G_\mathrm{cl}^{\ddagger}$ is the free energy
	barrier computed for a classical solvent.

\subsection{Model Systems}\label{sec:model_system}
We calculate rates for a series of bilinear vibronic coupling models of condensed phase electron transfer with potential
 \begin{equation}
 \label{eq:diab_pot}
     V(\hat{R}) = V_\mathrm{S} (\hat{s}) + V_\mathrm{B}(\hat{R})
 \end{equation}
 where the configuration vector $\hat{R} = \{\hat s,\hat Q\}$ represents the solvent polarization coordinate, $s$, and the bath coordinates, $Q$. 
 The diabatic potential energy matrix is 
 \begin{equation}
     V_\mathrm{S}(\hat{s}) = \left( \begin{array}{cc}
             A\hat{s}^2 + B\hat{s} + \varepsilon
    &  \Delta \\
    \Delta      & 
             A\hat{s}^2 - B\hat{s}
     \end{array} \right),
		\label{eq:potential_matrix}
 \end{equation}
 where $\varepsilon$ is the driving force of the reaction and $\Delta$ is the constant diabatic coupling.
 The solvent coordinate is linearly coupled to a thermal bath of $f$ harmonic oscillators,
 \begin{equation}
     V_\mathrm{B}(\hat{R}) = \sum_{j=1}^{f} \left[ \frac{1}{2} M_\mathrm{B} \omega_j^2 
				\left( \hat{Q}_j - \frac{c_j \hat{s}}{M_\mathrm{B} \omega_j^2} \right)^2 \right]
 \end{equation}
 where $M_B$ is the bath mass.  
 The bath is described by an Ohmic spectral
 density
 \begin{equation}
     J(\omega) = \eta \omega e^{-\frac{\omega}{\omega_c}}
 \end{equation}
 where $\omega_c$ is the cutoff frequency and $\eta$ is the dimensionless friction coefficient. 
 The spectral density is discretized into $f$ oscillators \cite{febCraig2005}
 \begin{equation}
     \omega_j = -\omega_\mathrm{c} \ln\left( \frac{j-0.5}{f} \right),
 \end{equation}
 with coupling strengths
 \begin{equation}
     c_j = \omega_j \left( \frac{2 \eta M_\mathrm{B} \omega_\mathrm{c}}{f \pi} \right)^{1/2}.
 \end{equation}
 The parameters for this model are shown in Table~\ref{tab:A_parameters}.
 
 \begin{table}[]
     \centering
     \caption{Linear vibronic coupling model parameters in atomic units unless otherwise indicated.}
     \label{tab:A_parameters}
    \singlespacing 
		\begin{tabular}{|c|c|}
     \hline
        Parameters   & Value \\
        \hline
        $A$  & $4.772 \times 10^{-3}$ \\
        $B$ & $2.288 \times 10^{-2}$ \\
        $\varepsilon$ & $0.0 - 0.2366$ \\
        $\Delta$ & $6.69 \times 10^{-7}$ - $1.20 \times 10^{-2}$\\
        $M_\mathrm{S}$ & $1836.0$ \\
        $M_\mathrm{B}$ & $1836.0$ \\
        $f$ & $12$ \\
        $\omega_\mathrm{c}$ & $2.28 \times 10^{-3}$ \\
        $\eta / M_{B} \omega_c$ & 1.0 \\
        $T$ & $150$ - $300$ K \\
        \hline
     \end{tabular}
 \end{table}

\color{black}
We further investigate a set of spin-boson models previously benchmarked using Heirarchy Equations of Motion (HEOM) 
against RPMD and Wolynes theory.~\cite{decLawrence2019} 
These models allow us to investigate the performance of MF-RPMD in the overdamped and underdamped regimes for both high frequency (B2 and B4) 
and low frequency (B1 and B3) systems. 
Parameters for all four models are shown in 
Table~\ref{tab:spin-boson_regimes}, with the 
shared parameters reported in Table~\ref{tab:B_parameters}. 

\begin{table}[]
     \centering
     \caption{Parameters (in atomic units) used for the diabatic potential in Eq.~\eqref{eq:diab_pot} that defines the spin-boson models B1-B4.}
     \label{tab:spin-boson_regimes}
     \singlespacing
        {\color{black}\begin{tabular}{|c|c|c|c|c|}
    \hline 
        Parameters & B1 & B2 & B3 & B4 \\
        \hline
        $A$& 0.125  &  8  & 0.125  & 8  \\
        $B$&  2.739  & 21.909  & 2.739  &  21.909 \\
        $\eta / M_{B} \omega_c$&  32  & 32  &  1 &  1 \\
        $\omega_\mathrm{c}$ & 0.5  & 4  &  0.5 & 4\\        
    \hline
        \end{tabular}}
\end{table}  
\begin{table}[]
     \centering
     \caption{Shared parameters for the spin-boson models B1-B4 in atomic units.}
     \label{tab:B_parameters}
    \singlespacing 
		{\color{black}\begin{tabular}{|c|c|}
     \hline
        Parameters   & Value \\
        \hline
        $\varepsilon$ & $15$ \\
        $\Delta$ & 0.1585, 1, 2.5119 \\
        $M_\mathrm{S}$ & 1.0 \\
        $M_\mathrm{B}$ & 1.0 \\
        $f$ & $12$ \\
        $\beta$ & 1 \\
        \hline
     \end{tabular}}
 \end{table}

 \color{black}
\subsection{Simulation Details}\label{sec:sim_details}
The numerator of the rate in Eq.~\ref{eq:kink-rate} 
	requires constrained sampling to the dividing surface 
    while the denominator requires sampling 
    the reactant region. We ensure sampling of all relevant configuration space using a windowing approach that introduces identity in the form of an integral over all solvent ring polymer centroid configurations. The resulting rate expression is
\textcolor{black}{
\begin{equation}\label{eq:rate-sample}
    k    = 
        \lim_{t \rightarrow t_\text{plateau}} \frac{ \int ds^\prime
        \langle \text{Tr}[\Gamma_\mathrm{k}] \delta(\Delta P_0) \dot{\Delta P}_0 h(\Delta P_t)\rangle_w}
        {\int ds^\prime \langle \text{Tr}[\Gamma_{1}] \rangle_w}
\end{equation}
}
where $\langle ... \rangle_w$ 
is used to indicate a phase space ensemble average over the nuclear bead configurations obtained 
by importance sampling from the constrained distribution
\begin{equation}
    w = e^{-\beta_N H_N^0\left( \left\{ R_\alpha, P_\alpha \right\} \right)- k_\mathrm{s}(\bar{s} - s')^2}, 
\end{equation}
where $\bar{s}$ is the solvent ring polymer centroid.
The integral in Eq.~\ref{eq:rate-sample} is evaluated  
performed by ensemble averaging for a series of windows 
centered at different values of $s'$, and then using the trapezoid rule. 
A standard Metropolis algorithm is used  
	to sample free ring polymer configurations
	from $e^{-\beta_N H_N^0\left( \left\{ R_\alpha,P_\alpha \right\} \right)}$ subject
 to a harmonic constraint with force constant $k=1000$ a.u. on the centroid.
We use 150 evenly spaced windows over a range of $s'$ values. 
\textcolor{black}{The initial value of $s'$ for each model is given in Table S1 in the Supporting Information.}

    All MF-RPMD simulations presented here are converged with respect to bead number, as reported in the results section, convergence numbers range from $N=16-32$.
    Initial conditions are sampled from $5\times 10^4$ decorrelated Monte Carlo
	steps for each window, 
	and time-evolved under the mean-field Hamiltonian in 
	Eq.~\ref{eq:mfrpmd_hamiltonian} using a fourth-order Adams-Bashforth-Moulton predictor-corrector integrator with a 
	time step of 0.05 a.u. \textcolor{black}{for models I-VIII and 0.0005 a.u. for models B1-4.} 
    The flux-side correlation function plateau in time is achieved between 500 a.u. and 3000 a.u. \textcolor{black}{for models I-VIII 
    and between 5 a.u. to 300 a.u. for the B1-4 models.}

The delta function at time zero in Eq.~\ref{eq:rate-sample} is approximated with the Gaussian,
\begin{equation}\label{eq:deltaGaus}
	\delta(\Delta P) \approx \frac{1}{\sigma\sqrt{2\pi}}\ \mathrm{e}^{-\frac{1}{2}\left(\frac{\Delta P}{\sigma}\right)^2}, 
\end{equation}
where $\sigma$ is treated a parameter controlling the width of the Gaussian. Results reported here employ $\sigma=0.01-0.001$; values that are sufficiently small to mimic a delta function, but sufficiently large to ensure meaningful statistics in terms of number of trajectories and number of windows contributing to the overall rate integral. We ensure that the rates reported are consistent across at least two smaller $\sigma$ values for each model. The analytical gradient of the population difference coordinate is used to obtain the initial velocity along the reaction coordinate in Eq.~\ref{eq:rate_expression_general}.

\textcolor{black}{The flux-side correlation function in Eq.~\ref{eq:kink-rate} assumes that the velocity is `positive' when the system 
is moving from reactant to product. In the inverted regime, the analytic time-derivative of the population difference 
coordinate can be negative for trajectories starting on the reactant side of the minimum, and
this can yield a reaction rate that is negative. We correct this by including a sign function that changes the sign of an 
individual bead's contribution to ensure that the bead-average velocity has a positive sign when the trajectory is moving towards product formation,}
\textcolor{black}{
\begin{align}
    \dot{\Delta P}&= \frac{1}{N}\sum_\alpha \dot{\Delta P}_\alpha\\
    \dot{\Delta P}_\alpha&=-\text{sgn}(\dot{\Delta P}_\alpha)
    \label{eq:sign_flip}
\end{align}
}

 \section{Results and Discussion} \label{sec:results}
\begin{figure}[h!]
     \centering
     \includegraphics[scale=1]{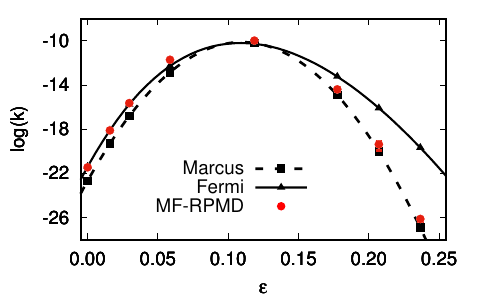}
 \caption{Nonadiabatic rates for bilinear coupling models I-VIII. 
 MF-RPMD rates with are shown with red circles. FGR rates (triangles and solid line) and the Marcus theory rates (squares and dashed line) are also shown.}
	\label{fig:drivingforce}
 \end{figure}

 We start with a study of MF-RPMD rates for nonadiabatic reactions
 with different driving forces at temperature $T=300$ K and with 
 the diabatic coupling  $\Delta = 6.69\times 10^{-7}$. 
 As shown in Fig.~\ref{fig:drivingforce}, we find 
 MF-RPMD rates in quantitative agreement 
 with FGR rates for Models I-IV that are in the 
 Marcus normal regime. The results shown in the figure are also 
 tabulated in Table~\ref{tab:drivingforce}, with all results calculated using $\sigma = 0.01$ in Eq.~\ref{eq:deltaGaus}. 
 \textcolor{black}{
 Note that we also calculate MF-RPMD rates using a centroid coordinate for normal regime models I, II, and IV, but do not plot the results since they are visually indistinguishable from the population coordinate results shown here. The rate expression with the centroid coordinate and the resulting MF-RPMD rates are provided in the SI.}
 
\begin{figure}[h!]
     \centering
     \includegraphics[scale=0.9]{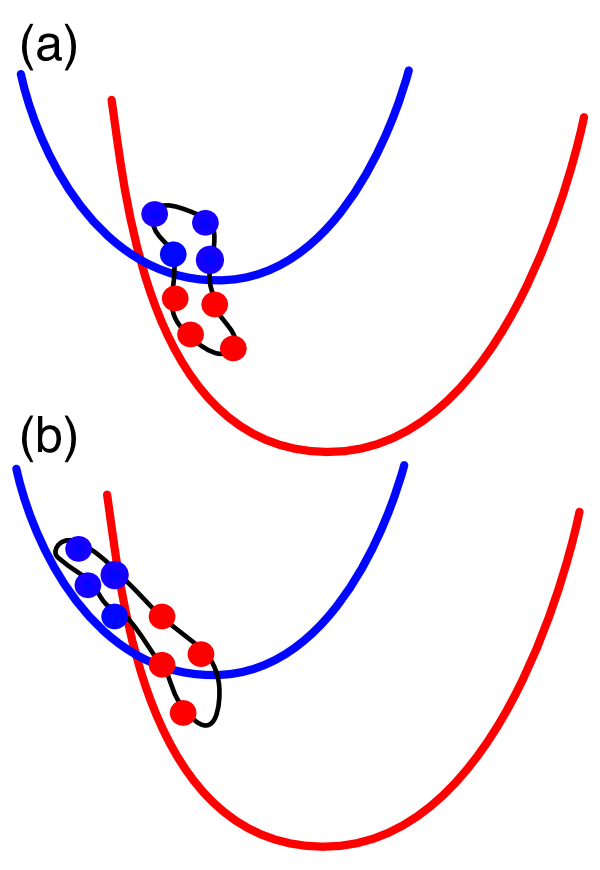}
 \caption{Cartoon of RP configurations in the inverted regime showing kinked configurations where some beads are in the reactant state and others are in the product state. Although configurations in (a) are quantum mechanically valid and necessary to capture solvent tunneling, in our MF-RPMD simulation they correspond to $\Delta P\neq 0$. Configurations in (b) do obey the $\Delta P=0$ criterion. For this reason, the MF-RPMD rates reported here fail to reproduce FGR rates in the inverted regime, instead yielding Marcus theory rates.}
	\label{fig:cartoon}
 \end{figure}

	\begin{table}[b]
		\centering
		\caption{Nonadiabatic reaction rates for models I-VIII}
		\label{tab:drivingforce}
		\singlespacing
		\begin{tabular}{|c|c|c|c|c|}
			\hline
			Model & $\varepsilon$ & log($k_\mathrm{MF}$) & log($k_\mathrm{FGR}$) & log($k_\mathrm{MT}$) \\
			\hline
			I & 0.00 & -21.42 & -21.37 & -22.65 \\
            II & 0.0158 & -18.09 & -18.08 & -19.30 \\
			III & 0.0296 & -15.60 & -15.73 & -16.79 \\
            IV & 0.0586 & -11.70 & -12.26 & -12.83 \\
			V & 0.1186 & -9.97 & -10.26 & -10.19 \\
			VI & 0.1776 & -14.38 & -13.22 & -14.91 \\
			VII & 0.2071 & -19.35 & -16.07 & -19.99 \\
            VIII & 0.2366 & -26.10 & -19.65 & -26.89 \\
			\hline
		\end{tabular}
	\end{table}
	In the inverted regime, for models V-VIII, 
 the MF-RPMD rate is in agreement with the Marcus Theory rate, despite the fact that MF-RPMD quantizes the solvent coordinate. We find that our definition of the dividing surface at $\Delta P=0$ is responsible for this discrepancy. As the population of each state is defined using the Boltzmann weight of that state, the state with the lower energy will have a higher population. In the normal regime, $\Delta P=0$ can be achieved by tunneling configurations where some beads are the left of the crossing with $\Delta P < 0$ and 
    some are to the right with $\Delta P > 0$ resulting in a bead-average value of zero. In
    the inverted regime, this remains true, however, tunneling configurations where all beads are 
    to the right of the crossing fail to meet the $\Delta P=0$ criterion as shown in the cartoon Fig.~\ref{fig:cartoon}.
\textcolor{black}{
This result highlights one of the limitations of using MF-RPMD: the lack of an explicit electronic variable means that constraining 
to kinked configurations places potentially unphysical constraints on the nuclear configurations.
In contrast, although KC-RPMD 
also employs a kink-constrained dividing surface, the constraint applies only to the auxiliary electronic variable introduced 
in that framework, allowing the method to capture nuclear tunneling in the Marcus inverted regime.
}
    
\begin{figure}[h!]
     \centering
     \includegraphics[scale=1]{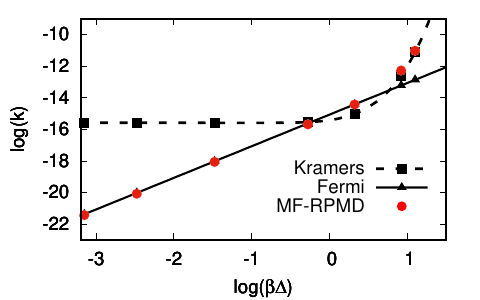}
 \caption{Reactions rates for model I with coupling strength modified to tune the system with varying coupling. MF-RPMD rates with the new population dividing surface are shown as red circles and the 
  error bars are within the symbol size. The FGR rate (triangles and solid line) and the Kramers 
 theory rate (squares and dashed line) are also shown}
	\label{fig:coupling}
 \end{figure}

One of the strengths of MF-RPMD is the ability to accurately
compute both adiabatic and nonadiabatic reaction rates within a single
framework. We demonstrate this using model I, a normal regime model at $T=300$ K, and we plot the results calculated using
$\sigma = 0.01$ in Eq.~\ref{eq:deltaGaus} against the Marcus 
and FGR rates in Fig.~\ref{fig:coupling}, and tabulate them in Table~\ref{tab:coupling}. We find that MF-RPMD rates move seamlessly
from good agreement with FGR rates in the nonadiabatic cases to good 
agreement with Kramer's theory rates in the adiabatic regime.

	\begin{table}[]
		\centering
		\caption{Reaction rates for model I with variable coupling}
		\label{tab:coupling}
		\singlespacing
		\begin{tabular}{|c|c|c|c|c|c|}
			\hline
			$\Delta$ & $N$ & log($k_\mathrm{MF}$) & log($k_\mathrm{FGR}$) & log($k_\mathrm{KT}$) \\
			\hline
			 $6.69 \times 10^{-7}$ & 16 & -21.42 & -21.37 & -15.57 \\
			 $3.16 \times 10^{-6}$ & 16 & -20.07 & -20.02 & -15.58 \\
			 $3.16 \times 10^{-5}$ & 16 & -18.06 & -18.02 & -15.58 \\
			 $5.01 \times 10^{-4}$ & 32 & -15.68 & -15.62 & -15.54 \\
			 $2.00 \times 10^{-3}$ & 32 & -14.42 & -14.42 & -15.02 \\
			 $7.94 \times 10^{-3}$ & 32 & -12.27 & -13.22 & -12.60 \\
			 $1.20 \times 10^{-2}$ & 32 & -11.02 & -12.86 & -11.11 \\
			\hline
		\end{tabular}
	\end{table} 

\begin{figure}[h!]
     \centering
     \includegraphics[scale=1]{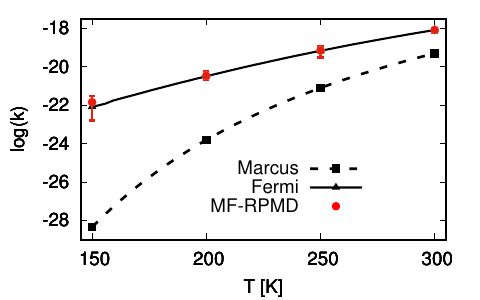}
 \caption{ET rates as function of temperature for model II with $\varepsilon=0.0158$. 
    MF-RPMD rates with the new population dividing surface are shown as
	red circles. The Fermi's golden rule (triangles and solid line) and the Kramers 
 theory (squares and dashed line) rates are also shown}
	\label{fig:temperature}
 \end{figure}

    Further, we examine the effect of temperature on the rate constant. For this we choose a model in
    the normal regime, model II, in the weak coupling limit $\Delta = 6.69\times 10^{-7}$.
    For these simulations we use a value of $\sigma = 0.001$ in Eq.~\ref{eq:deltaGaus}. The results 
    are shown in Fig.~\ref{fig:temperature} and tabulated in Table~\ref{tab:temperature}. Consistent with
    our other results in the normal regime, we achieve quantitative agreement with the rates from Fermi's golden
    rule. As expected, we do have to increase the number of beads from 16 to 24 to get converged results at
    lower temperatures. 
    \textcolor{black}{
    At very high temperatures, as suggested by earlier studies,~\cite{lawrence2020path} the present rate expression is 
    expected to perform poorly given the {\it adhoc} nature of the dividing surface introduced. Indeed, we find 
    this is the case at $T=5000K$ comparing the 8-bead MF-RPMD rate (1-bead is the zero-coupling limit) against the corresponding
    FGR rate, we find the rate differs by almost an order of magnitude, $k_\text{FGR}=5.36\times 10^{-12}$ a.u. and $k_\text{MF}=1.56\times 10^{-13}$  a.u. }
    
 \begin{table}[b]
		\centering
		\caption{MF-RPMD rates for the linear vibronic coupling model with varying temperature}
		\label{tab:temperature}
		\singlespacing
		\begin{tabular}{|c|c|c|c|c|}
			\hline
			T (K) & $N$ & log($k_\mathrm{MF}$) & log($k_\mathrm{FGR}$) & log($k_\mathrm{MT}$) \\
			\hline
			150 & 24 & -21.84 & -22.07 & -28.33 \\
			200 & 24 & -20.46 & -20.48 & -23.80 \\
            250 & 16 & -19.15 & -19.16 & -21.09 \\
			300 & 16 & -18.07 & -18.08 & -19.30 \\
			\hline
		\end{tabular}
	\end{table}

\begin{figure}[h!]
\centering
\includegraphics[scale=0.4]{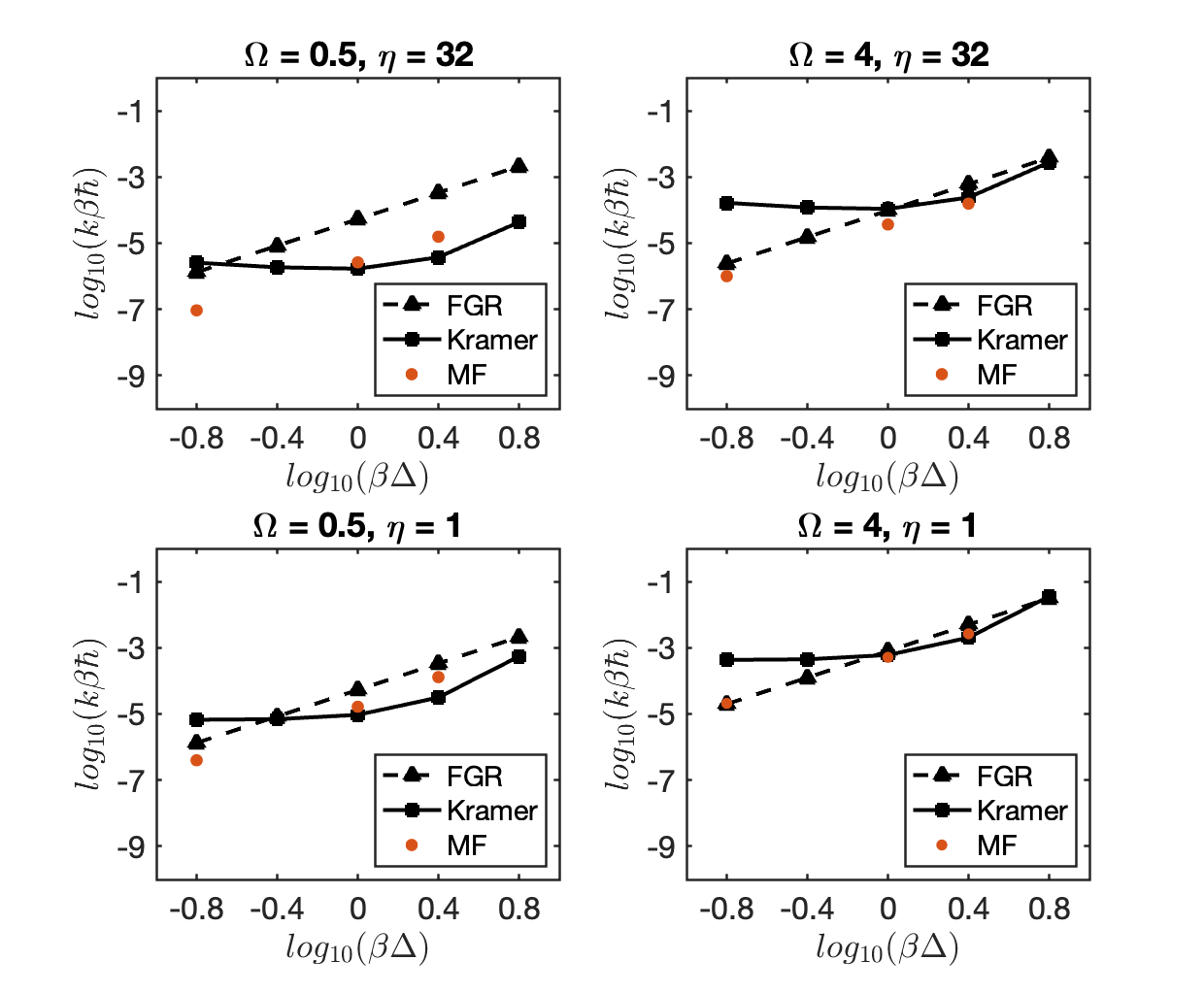}
 \caption{\textcolor{black}{MF-RPMD reaction rates for the spin-boson models B1-B4 as a function of coupling strength. The MF-RPMD rates (red circles) are compared against with The Fermi's golden rule (triangles and solid line) and the Kramers 
 theory (squares and dashed line)}}
\label{fig:Spin-boson-results}
\end{figure}

\textcolor{black}{
    Finally, we present the MF-RPMD rates for models B1-B4. The results 
    are shown in Fig.~\ref{fig:Spin-boson-results} and tabulated in Table~\ref{tab:spin-boson}. Overall, we find relatively good agreement when against FGR and Kramer's rate theory results as the systems move from the nonadiabatic to the adiabatic regime. The most significant error is for the case of low frequency
    and strong system-bath coupling where we find the nonadiabatic 
    MF-RPMD rates significantly underestimate the FGR rates. 
    This appears to be in keeping with the limitations of the current rate
    expression that does not correctly capture classical-limit behavior.
    }

\begin{table}[h]
		\centering
		\caption{MF-RPMD rates for the spin-boson models B1-B4}
		\label{tab:spin-boson}
		\singlespacing
		\begin{tabular}{|c|c|c|c|c|c|}
		\hline
		Model & log($\beta\Delta$) & log($k_\mathrm{MF}$) & log($k_\mathrm{FGR}$) & log($k_\mathrm{Kramer})$ \\
		\hline
		\multirow{3}{*}{B1} & -0.8 & -7.03 & -5.88 & -5.59  \\
			   & 0 & -5.58 & -4.28 & -5.77\\
               & 0.4 & -4.80 & -3.48& -5.43 \\
               \hline
		\multirow{3}{*}{B2}  & -0.8 & -6.00  & -5.62 & -3.78 \\
			   & 0 & -4.44 & -4.02 & -3.96 \\
               & 0.4 & -3.81 & -3.22 & -3.62 \\
               \hline
        \multirow{3}{*}{B3}  & -0.8 & -6.40 &  -5.87 &-5.18 \\
			   & 0 & -4.78 & -4.27 & -5.03   \\
               & 0.4 & -3.89 & -3.47 & -4.50   \\
               \hline
        \multirow{3}{*}{B4}  & -0.8 &-4.68  & -4.70 & -3.37 \\
			   & 0 & -3.29  & -3.10 & -3.22  \\
               & 0.4 & -2.58 & -2.30 & -2.70 \\
               \hline
		\end{tabular}
	\end{table}

 \section{Conclusions} \label{sec:conclusion}
 We introduce a new population difference reaction coordinate and a kink-constrained dividing surface for the calculation of rates using MF-RPMD.
 This new rate formulation represents a simplification of the previously introduced MF-RPMD rate expression for multi-level systems with an implementation that is designed to enable atomistic system simulations. 
 We have shown that the MF-RPMD rate expression introduced here is accurate over a wide range of electronic coupling strengths, captures nonadiabatic rates accurately over a range of driving forces, and is able to capture nuclear tunneling effects at low temperatures.

\textcolor{black}{
Future directions include identifying modifications to the current protocol to enable capturing tunneling in the inverted regime, and addressing the inaccuracies in rate as we approach classical-limit behavior. 
}
While the current implementation is scalable to atomistic simulations where reactant and product state are defined in the diabatic framework using methods like the multi-state empirical valence band model,~\cite{febVoth2006} we are also investigating working with the adiabatic rather than diabatic states to enable on-the-fly simulations, an attractive direction given that the MF-RPMD Hamiltonian is agnostic to the choice of electronic state representation.

\section{Acknowledgements} \label{sec:acknowledgements}
        N.A., B.A.J., and N. L. acknowledge support from the U.S. Department of Energy, Office of Basic Energy Sciences, Division of Chemical Sciences, Geosciences and Biosciences under Award DE-FG02-12ER16362 (Nanoporous Materials Genome: Methods and Software to Optimize Gas Storage, Separations, and Catalysis). B.A.J. acknowledges current support by the DOE Office of Science, Office of Basic Energy Sciences, Division of Chemical Sciences, Geosciences, and Biosciences, Condensed Phase and Interfacial Molecular Science program. S.B. acknowledges support from Cornell University, Department of Chemistry and Chemical Biology.

\begin{suppinfo}
%
%
Additional simulations details for MF-RPMD implementation and details of the analytic expression for calculating the FGR rate for bilinear coupling model Hamiltonians are provided in the supporting information.
%
\end{suppinfo}

\bibliography{mfrpmd-paper}

\end{document}